\documentclass[preprint,amsmath,amssymb,prd,superscriptaddress]{revtex4}
\usepackage[dvips]{graphicx}
\usepackage{color}
\usepackage{multirow}
\usepackage{array}
\usepackage{float}
\usepackage{amsmath,amssymb,amsfonts,amstext,amsthm}
\usepackage{algorithm,algorithmic}
\newcommand{\p}{\partial}
\newcommand{\pslash}{p\kern-1ex /}
\newcommand{\lslash}{l\kern-1ex /}
\newcommand{\kslash}{k\kern-1ex /}
\newcommand{\dslash}{\p\kern-1.2ex /}
\newcommand{\Dslash}{{\cal D}\kern-1.5ex /}
\newcommand{\Tr}{{\rm Tr}}
\newcommand{\tr}{{\rm tr}}
\newcommand{\re}{{\rm Re}}

\newcommand{\bea}{\begin{eqnarray}}
\newcommand{\eea}{\end{eqnarray}}

\newcommand{\BAN}{\begin{eqnarray*}}
\newcommand{\EAN}{\end{eqnarray*}}
\begin{document}

\newcommand{\NTU}{
  Department of Physics,
  National Taiwan University, Taipei~10617, Taiwan
}

\newcommand{\CQSE}{
  Center for Quantum Science and Engineering,
  National Taiwan University, Taipei~10617, Taiwan
}

\newcommand{\CTS}{
  Center for Theoretical Sciences,
  National Taiwan University, Taipei~10617, Taiwan
}

\preprint{NTUTH-16-505A}

\title{The $\beta$-function of $SU(3)$ gauge theory with $ N_f = 10 $ 
       massless fermions in the fundamental representation}

\author{Ting-Wai~Chiu}
\affiliation{\NTU}
\affiliation{\CQSE}
%\affiliation{\CTS}

\begin{abstract} 
We present the first study of the discrete $\beta$-function of $ SU(3) $ lattice  
gauge theory with 10 massless domain-wall fermions in the fundamental representation. 
The renormalized coupling is obtained by the finite-volume gradient flow scheme, 
and the discrete $\beta$-function is extrapolated to the continuum limit by the step-scaling method. 
Our result of the discrete $\beta$-function (with $ s = 2 $) suggests that
this theory possesses an infrared fixed point around $ g_c^2 \sim 7.0 $ for $ c = \sqrt{8t}/L = 0.3 $.

\end{abstract}

\maketitle

%\pacs{11.15.Ha,11.30.Rd,12.38.Gc}

\section{Introduction}

In weak-coupling perturbation theory (WCPT), the $\beta$-function of non-abelian gauge theory 
with $ N_f $ massless fermion can possess a non-trivial infrared fixed-point (IRFP), 
besides the ultraviolet fixed point (UVFP) at $ g^2 = 0 $, 
provided that $ N_f $ is within a range (i.e., the conformal window), and   
these theories are infrared conformal \cite{Caswell:1974gg, Banks:1981nn}. 
The relevance of these infrared conformal theories 
to high energy phenomenology is the possibility that the Higgs scalar in the Standard Model (SM) 
might be a bound state of fermion-antifermion in a new non-abelian gauge theory with 
$ N_f $ massless fermions just below the edge of the conformal window, as an approximate Nambu-Goldstone boson 
resulting from breaking the conformal symmetry. In these theories, 
unlike QCD with spontaneously broken chiral symmetry, 
the chiral symmetry is unbroken, thus the scalar might emerge as the lightest bound state 
rather than the pseudoscalar, analogous to the scenario depicted in Ref. \cite{Yamawaki:1985zg} 
which proposed a plausible resolution to the problem of flavor-changing neutral-current 
in the technicolor model \cite{Weinberg:1975gm,Susskind:1978ms}.
This is one of the basic motivations of studying the $\beta$-function and 
the conformal window of non-abelian gauge theory with $ N_f $ massless fermions.  

In general, the conformal window depends on the gauge group as well as the representation of the massless fermions. 
For $ SU(3) $ gauge theory with $ N_f $ massless fermions in the fundamental representation, 
WCPT to 2-loop order gives the (approximate) conformal window $ 8 < N_f \le 16 $ \cite{Caswell:1974gg}. 
For $ N_f = 10 $, its IRFP is around $ g^2 \sim 28 $.
Obviously, WCPT is not supposed to give reliable answers at such strong coupling. 
This calls for nonperturabtive study of the infrared behavior 
of the running coupling of nonabelian gauge theory with $ N_f $ massless fermions. 
This is a fundamental problem in quantum field theory, regardless of whether the Higgs scalar 
is a composite scalar arising from the breaking of the conformal symmetry or not. 

Lattice gauge theory provides a viable framework for nonperturbative study of vector gauge theories. 
Since we are dealing with massless fermions, it is vital to use lattice fermions with exact chiral symmetry,
i.e., domain-wall \cite{Kaplan:1992bt} /overlap \cite{Neuberger:1997fp, Narayanan:1994gw} fermions, 
having exactly the same flavor symmetry as their counterparts in the continuum. 
In this paper, we focus on the $ SU(3) $ lattice gauge theory with 
10 massless domain-wall fermions (DWF) in the fundamental representation. 
To preserve the chiral symmetry maximally on a lattice with finite extension  
in the fifth dimension, we use the optimal DWF with the $ R_5 $ symmetry \cite{Chiu:2015sea},  
which has the effective 4-dimensional lattice Dirac operator exactly 
equal to the ``shifted" Zolotarev optimal rational approximation 
of the overlap operator, with the approximate sign function $ S(H) $ 
satisfying the bound $ 0 \le 1-S(\lambda) \le 2 d_Z $ 
for $ \lambda^2 \in [\lambda_{min}^2, \lambda_{max}^2] $,
where $ d_Z $ is the maximum deviation $ | 1- \sqrt{x} R_Z(x) |_{\rm max} $ of the
Zolotarev optimal rational polynomial $ R_Z(x) $ of $ 1/\sqrt{x} $ 
for $ x \in [1, \lambda_{max}^2/\lambda_{min}^2] $, with degrees $ (n-1,n) $ for $ N_s = 2n $.

To obtain the renormalized coupling of gauge theory on a finite lattice with volume $ L^4 $,   
we use the finite-volume gradient flow scheme \cite{Fodor:2012td}, 
which is based on the idea of continuous-smearing \cite{Narayanan:2006rf} 
or equivalently the gradient flow \cite{Luscher:2010iy}
to evaluate the expectation value $ t^2 \langle E \rangle $, 
where $ E $ is the energy density of the gauge field, and $ t $ is the flow time.
This amounts to solving the discretized form of the following equation 
\BAN
\frac{d B_\mu}{dt} = D_\nu G_{\nu \mu},  
\EAN 
with the initial condition $ B_\mu |_{t=0} = A_{\mu} $, 
where $ G_{\nu\mu} = \partial_\nu B_\mu - \partial_\mu B_\nu + [B_\nu, B_\mu] $, and 
$ D_\nu G_{\nu\mu} = \partial_\nu G_{\nu\mu} + [ B_\nu, G_{\nu\mu} ] $.
As shown in Ref. \cite{Luscher:2010iy}, 
the gradient flow is a process of averaging gauge field over a spherical region of root-mean-square 
radius $ R_{rms} = \sqrt{8 t} $. Moreover, since $ t^2 \langle E \rangle $ is proportional to the 
renormalized coupling, one can use $ c = \sqrt{8t}/L $ as a constant to define a renormalization scheme 
on a finite lattice, and obtain
\bea
\label{eq:g2L}
g_c^2(L,a) = \frac{16 \pi^2}{3[1+\delta(c,a/L)]} \langle t^2 E(t) \rangle,  
\hspace{10mm}  E(t) = \frac{1}{2} F_{\mu\nu} F_{\mu\nu}(t),
\eea
where $ a $ is the lattice spacing depending on the bare coupling $ g_0 $, $ E $ is the energy density, 
and the numerical factor on the RHS of (\ref{eq:g2L}) is fixed 
such that $ g_c^2(L,a) = g^2_{\overline{\text{MS}}} $ to the leading order.
Here the coefficient $ \delta(c,a/L) $ includes the tree-level finite volume 
and finite lattice spacing corrections \cite{Fodor:2014cpa}. 
In this paper, we use the Wilson flow, the Wilson action, and the clover observable, 
the so called WWC scheme, which is known to have very small tree-level cutoff effects \cite{Fodor:2014cpa}.
Moreover, we fix $ c = \sqrt{8t}/L = 0.30 $. 
 
For any input value of $ g^2(L,a) = g^2$, we compute the discrete $\beta$-function (at finite $ a $)    
\bea
\label{eq:DBF}
\beta(s, a/L, g^2) = -\frac{g^2(sL,a) - g^2(L,a)}{\ln(s^2)}, 
\eea 
for all lattice pairs $ (L, sL) $ with fixed $ s $. Assuming the discretization error of 
$ \beta(s, a/L, g^2) $ behaves as $ O(a^2) $, one can extrapolate $ \beta(s, a/L, g^2) $ 
to the continuum limit, i.e., $ \lim_{a\to 0} \beta(s, a/L, g^2) = \beta(s, g^2) $, 
the so-called step-scaling method \cite{Luscher:1991wu}.
Moreover, if $ \beta(s,g^2) $ is determined for several values of $ s $,  
then it can be extrapolated to $ s = 1 $,  
\bea 
\lim_{s \to 1} \lim_{a \to 0} \beta(s, a/L,g^2) = \lim_{s \to 1} \beta(s, g^2) = \beta(g^2) 
= \frac{d g^2}{d \ln \mu^2},    
\eea
which corresponds to the conventional $\beta$-function in the continuum. 
If $ \beta(g^2) $ has an IRFP, then $ \beta(s, g^2) $ also has a corresponding IRFP, and vice versa.
In this paper, we determine the discrete $\beta$-function (with $ s = 2 $) of $ SU(3) $ lattice gauge theory 
with 10 massless optimal domain-wall fermions in the fundamental representaion, 
using three lattice pairs $ (L, 2L)/a = (8, 16) $, (10, 20), and (12, 24) 
for extrapolation to the continuum limit $ a \to 0 $. 

%To fix our notation, we recall the $\beta$-function to the 2-loop order in the $ SU(3) $ 
%gauge theory with $ N_f $ massless fermions in the fundamental representation, 
%\bea
%\label{eq:beta_2loop}
%\beta(g^2) = \frac{d g^2}{d \ln \mu^2} = - \frac{b_1}{(4 \pi)^2} g^4 - \frac{b_2}{(4 \pi)^4} g^6 + O(g^8),   
%\eea
%where $ b_1 = 11 - 2 N_f/3 $, and $ b_2 = 102 - 38 N_f/3 $.

At this point, we briefly summarize previous studies on the $ SU(3) $ gauge theory with $ N_f = 10 $ fermions. 
Using the Wilson fermion and the plaquette gauge action, Hayakawa et al. \cite{Hayakawa:2010yn}
computed the discrete $\beta$-function with the renormalized coupling in the Schr\"odinger functional scheme, 
and observed an IRFP at strong coupling. 
Since the Wilson fermion breaks the chiral symmetry explicitly, it is unclear to what extent the lattice 
artifact affects their results.  
Instead of computing the discrete $\beta$-function, Appelquist et al. \cite{Appelquist:2012nz} 
studied the low-lying (pseudoscalar, vector, and axial-vector mesons) spectrum of 
the $ SU(3) $ gauge theory with $ N_f = 10 $ light domain-wall fermions,  
with results consistent with infrared conformality for the fermion mass $ m a \ge 0.010 $. However, their 
results cannot rule out the possibility that the theory might undergo spontaneously chiral symmetry breaking 
at smaller fermion mass.  

%The outline of this paper is as follows.    

\section{Simulations}

For the gauge action, we use the Wilson plaquette gauge action 
\BAN
S_g(U) = \frac{6}{g_0^2} \sum_{plaq.}\left\{1-\frac{1}{3} \re \Tr (U_p) \right\}, 
\hspace{4mm} \beta = \frac{6}{g_0^2},  
\EAN
where $ g_0 $ is the bare coupling.
For the massless fermions, we use the optimal DWF with $ R_5 $ symmetry \cite{Chiu:2015sea}. 
The action for one-flavor optimal DWF can be written as   
\bea
\label{eq:S_odwf}
S(\bar\Psi, \Psi, U) = \bar\Psi_{x,s} \left[
  (\omega_s D_w + 1)_{xx'} \delta_{ss'}
 +(\omega_s D_w - 1)_{xx'} L_{ss'} \right] \Psi_{x',s'},
\eea
where the indices $ x $ and $ x' $ denote the sites on the 4-dimensional space-time lattice,
and $ s $ and $ s' $ the indices in the fifth dimension, while
the lattice spacing $ a $ and the Dirac and color indices have been suppressed.
Here $ D_w $ is the standard Wilson-Dirac operator minus the parameter $ m_0 \in (0,2) $.  
The operator $ L $ is independent of the gauge field, and it can be written as  
\BAN
L = P_+ L_+ + P_- L_-, \quad P_\pm = (1\pm \gamma_5)/2,
\EAN
and
\bea
\label{eq:L}
(L_+)_{ss'} = (L_-)_{s's}= \left\{ 
    \begin{array}{ll} 
      - m_q/(2m_0) \delta_{N_s,s'}, & s = 1, \\  
      \delta_{s-1,s'}, & 1 < s \leq N_s,   
    \end{array}\right.
\eea
where $ N_s $ is the number of sites in the fifth dimension,    
$m_q $ is the bare fermion mass, and $ m_0 \in (0, 2) $. 
For massless DWF, $ m_q $ is set to zero.  
Besides (\ref{eq:S_odwf}),    
the action for the Pauli-Villars fields with $ m_q = 2 m_0 $ has to be included   
for the cancellation of the bulk modes, which is exactly the same as (\ref{eq:S_odwf}) 
except for $ m_q = 2 m_0 $ in $ L_\pm $ (\ref{eq:L}). 
Thus the action for $ SU(3) $ lattice gauge theory with 10 massless optimal DWF can be written as 
\BAN
S_g(U) + \sum_{f=1}^{10} \left\{ S_{m_q=0}( \bar\Psi, \Psi, U)_f + S^{PV}_{m_q=2 m_0} (\bar\Phi, \Phi, U)_f \right\}. 
\EAN
In this paper, we set $ m_0 = 1.8 $, and $ N_s = 16 $.
The optimal weights $ \omega_s $ are obtained by the formula given in \cite{Chiu:2015sea}, 
with $ \lambda_{min} = 0.05 $ and $ \lambda_{max} = 6.2 $. 

Following the procedures of even-odd preconditioning and the Schur decomposition given in Ref. \cite{Chiu:2013aaa}, 
the partition function for the $ SU(3) $ gauge theory with $ N_f = 10 $ massless optimal DWF can be written as 
\bea
\label{eq:Z_nf10}
Z = \int[dU]\prod_{i=1}^5 [d\phi^{\dag}]_i[d\phi]_i 
\exp \left(-S_g[U]- \sum_{i=1}^5 \phi^\dagger_i (C_{PV}^\dagger)_i ( C C^\dagger)^{-1}_i (C_{PV})_i \phi_i \right),
\eea  
where $ \phi_i $ and $ \phi^\dagger_i $ are pseudofermion fields, and  
\BAN
\label{eq:C_def}
C &=& 1 - M_5 D_w^{\text{OE}} M_5 D_w^{\text{EO}}, \\
\label{eq:m5}
M_5 &=& = \left\{(4-m_0) + \omega^{-1/2}_s [(1-L)(1+L)]^{-1}_{s,s'} \omega^{-1/2}_{s'} \right\}^{-1}.
\EAN

For the $ 24^4 $ lattice at strong couplings $ \beta = 6/g_0^2 = 6.50 $ and 6.45,
a novel $ N_f = 2 $ pseudofermion action is used for the HMC simulations, 
which turns out to be more efficient than that in (\ref{eq:Z_nf10}). 
This $ N_f = 2 $ pseudofermion action is based on the exact pseudofermion action 
for one-flavor DWF \cite{Chen:2014hyy}. 
Following the notations and formulas in Ref. \cite{Chen:2014hyy}, 
this novel $ N_f = 2 $ pseudofermion action can be written as $ S = \phi^\dagger K(m)^\dagger K(m) \phi $, where 
\BAN
K(m) = 1 +k \gamma_5 v^{T}\omega^{-1/2}\frac{1}{H_T(m)}\omega^{-1/2}v, \hspace{4mm}
v =
\left(
\begin{array}{cc}
   v_+  &          0 \\
   0       &     v_-
\end{array}
\right)_{Dirac}.
\EAN
Then the partition function for the $ SU(3) $ gauge theory with $ N_f = 10 $ massless optimal DWF can be written as 
\bea
\label{eq:Z1_nf10}
Z' = \int[dU]\prod_{i=1}^5 [d\phi^{\dag}]_i[d\phi]_i 
\exp \left(-S_g[U]- \sum_{i=1}^5 \phi^\dagger_i K(0)^\dagger K(0) \phi_i \right).
\eea  

We perform HMC simulations with (\ref{eq:Z_nf10}) or (\ref{eq:Z1_nf10}) 
on the 5-dimensional lattice $ L^4 \times 16 $, 
for lattice sizes $ L/a = 8, 10, 12, 16, 20, 24$, 
each with 12 bare coulings ($ g_0 $) parametrized 
by $ \beta = 6/g_0^2 = 15.0, 12.0, 10.0, 9.0, 8.0, 7.5, 7.0, 6.8, 6.7, 6.6, 6.5, 6.45 $. 
Thus we have a total of 72 gauge ensembles. The boundary conditions of the gauge field 
are periodic in all directions, while the boundary conditions of the pseudofermion fields
are antiperiodic in all directions. The simulations are performed on GPU clusters 
at National Taiwan University, which are dedicated to lattice gauge theory.
All gauge ensembles (except for $24^4$ at $ \beta = 6.45 $, 6.50) are 
simulated in one single stream, with one GPU or two GPUs in one computing node. 
%For simulations with multi-GPUs, we take advantage of the salient feature Peer-to-Peer (P2P) of Nvidia GPUs 
%with hardware architecture greater than or equal to sm\_20,  
%e.g., TITAN-X (sm\_52), TITAN-Z (sm\_35), GTX970 (sm\_52), C2070 (sm\_20), etc. 
%Then the speed up of the conjugate gradient with multi-GPUs can attain 3.6 for 4 GPUs on the same motherboard.     
%It takes about 14 months to complete all simulations of 72 gauge ensembles.
For each gauge ensemble, we generate 3000-5000 trajectories after thermalization, 
and sample one configuration every 5 trajectories, which yields 600-1000 configurations for measurements.  
%The statistical error of $ t^2 \langle E \rangle $ is estimated by the jackknife method 
%with a bin size of 10-15 configurations of which the statistical error saturates. 
%For any gauge ensemble, the statistical error of 
%$ t^2 \langle E \rangle $ is less than 0.5\% for $ c = \sqrt{8t}/L = 0.3 $. 

The chiral symmetry breaking due to finite $ N_s = 16 $ can be measured in terms of 
the residual mass of the massless fermion \cite{Chen:2012jya}, 
\BAN
\label{eq:mres}
m_{res} = \frac{\left< \tr (D_c^{-1})_{0,0} \right>_U}
               {\left< \tr[\gamma_5 {D_c} \gamma_5 {D_c}]^{-1}_{0,0} \right>_U},
\EAN
where $ D_c^{-1} $ denotes the fermion propagator,
tr denotes the trace running over the color and Dirac indices, 
and the brackets $ \left< \cdots \right>_U $ denote averaging over all configurations of the gauge ensemble. 
The residual mass is less than $ 5 \times 10^{-5} $ for any gauge ensemble in this work. 

\begin{figure}[htb]
\begin{center}
\includegraphics*[width=10cm,clip=true]{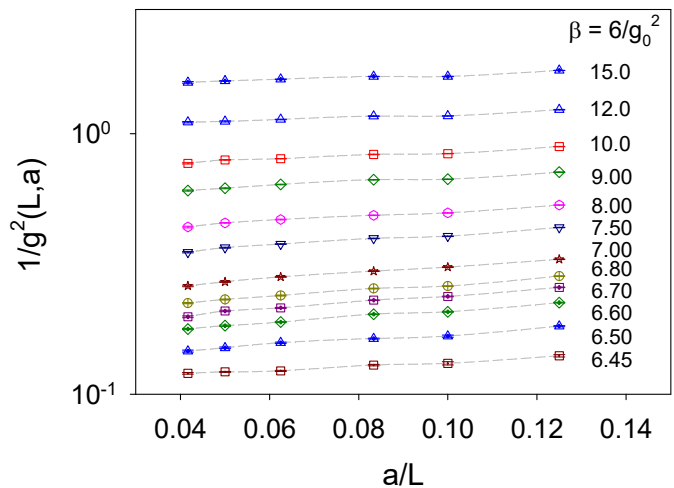}
\caption{The data of $ 1/g^2(L,a) $ versus $ a/L $, for $ L/a = 8, 10, 12, 16, 20, 24 $, and 
for 12 values of $ \beta = 6/g_0^2 $ (as shown at the RHS of the figure). 
The data points of the same $ \beta = 6/g_0^2 $ are connected by a dotted line. 
The $y$-axis is in the common logarithm scale.}
\label{fig:Ig2L_Nf10}
\end{center}
\end{figure}

\section{Results}

Fixing $ c = \sqrt{8t}/L = 0.3 $, we measure the renormalized coupling $ g^2(L,a) $ according to (\ref{eq:g2L}). 
The statistical error of $ t^2 \langle E \rangle $ is estimated by the jackknife method 
with a bin size of 10-15 configurations of which the statistical error saturates. 
For each gauge ensemble, the statistical error of $ g^2(L,a) $ is less than $ 0.5\% $. 
In Fig. \ref{fig:Ig2L_Nf10}, we plot the data of $ 1/g^2(L,a) $ versus $ a/L $, 
for 12 different values of $ \beta = 6/g_0^2 $. Here the $y$-axis is in the common logarithm scale.
The data points with the same $ \beta $ are connected by a dotted line. 
For each $ \beta $, $ 1/g^2(L,a) $ is a monotonic increasing function of $ a/L $.  

To compute the discrete $\beta$-function (\ref{eq:DBF}) for any $ g^2(L,a) $, 
it requires $ g^2(L,a) $ and $ g^2(2L,a) $ at the same $ \beta=6/g_0^2 $ which 
in general is not equal to the 12 values of $ \beta = 6/g_0^2 $ in our simulations.
In other words, the value of $ g^2(L,a) $ at any $ \beta \in [6.45, 15.0] $ 
has to be obtained by interpolation, based on the 12 data points of $ g^2(L,a) $ for each $ L $.
To this end, we use the cubic spline interpolation to obtain $ g^2(L,a) $ for any $ \beta \in [6.45, 15.0] $. 
Then, for any input value of $ g^2(L) $, 
the discrete $\beta$-function (\ref{eq:DBF}) can be obtained for any lattice pair $ (L, 2L) $. 
Since both the action and the observable only contain $ O(a^2) $ corrections, 
the discrete $\beta$-function $ \beta(2, a/L, g^2) $ can be linearly extrapolated to the continuum limit 
as a function of $ (a/L)^2 $, using 3 data points of different lattice spacings corresponding to 
the lattice pairs: (8, 16), (10, 20) and (12, 24).  

\begin{figure}[H]
\begin{center}
\begin{tabular}{@{}cccc@{}}
\includegraphics*[height=5cm,width=7.5cm,clip=true]{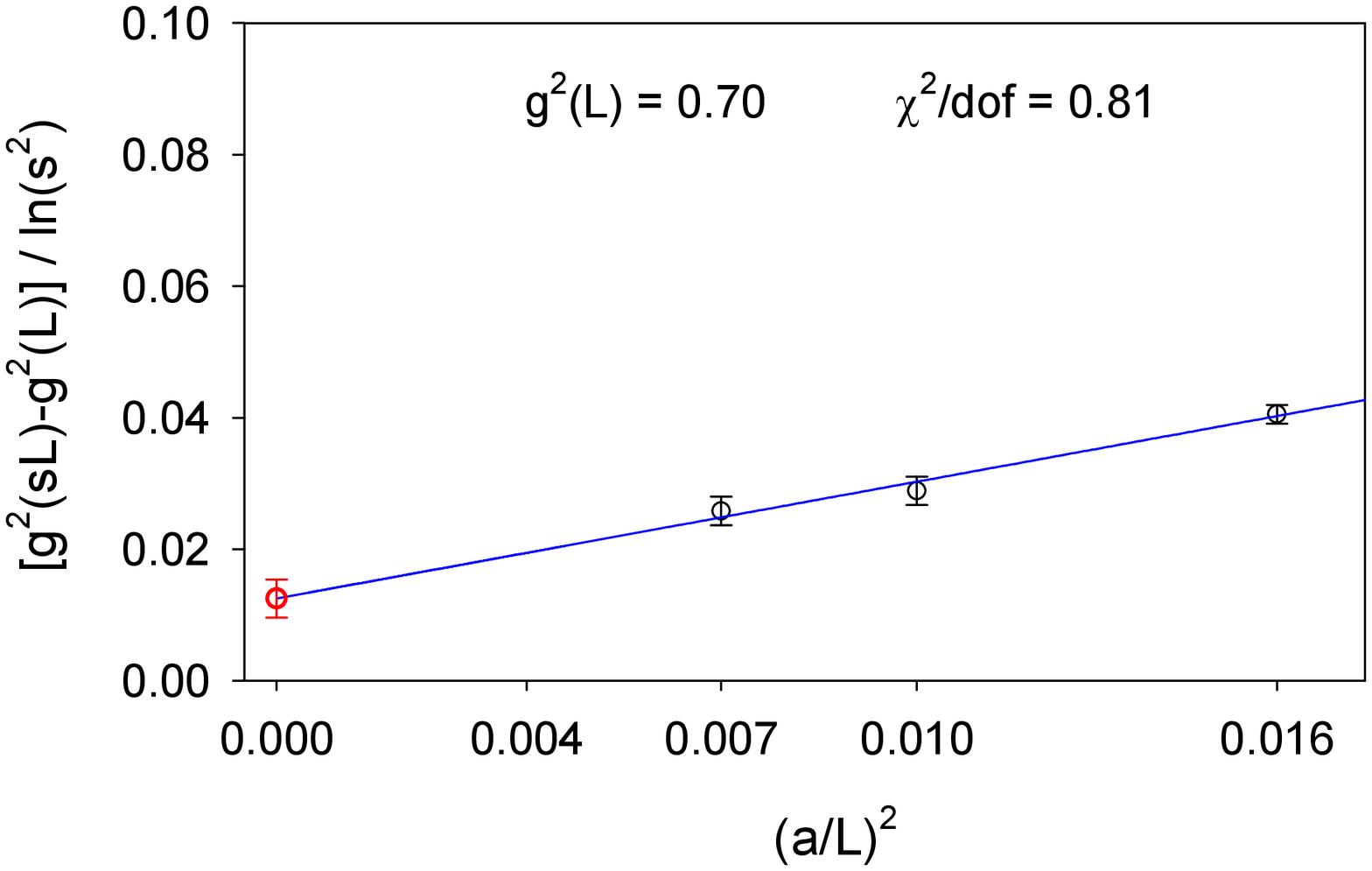}
&
\includegraphics*[height=5cm,width=7.5cm,clip=true]{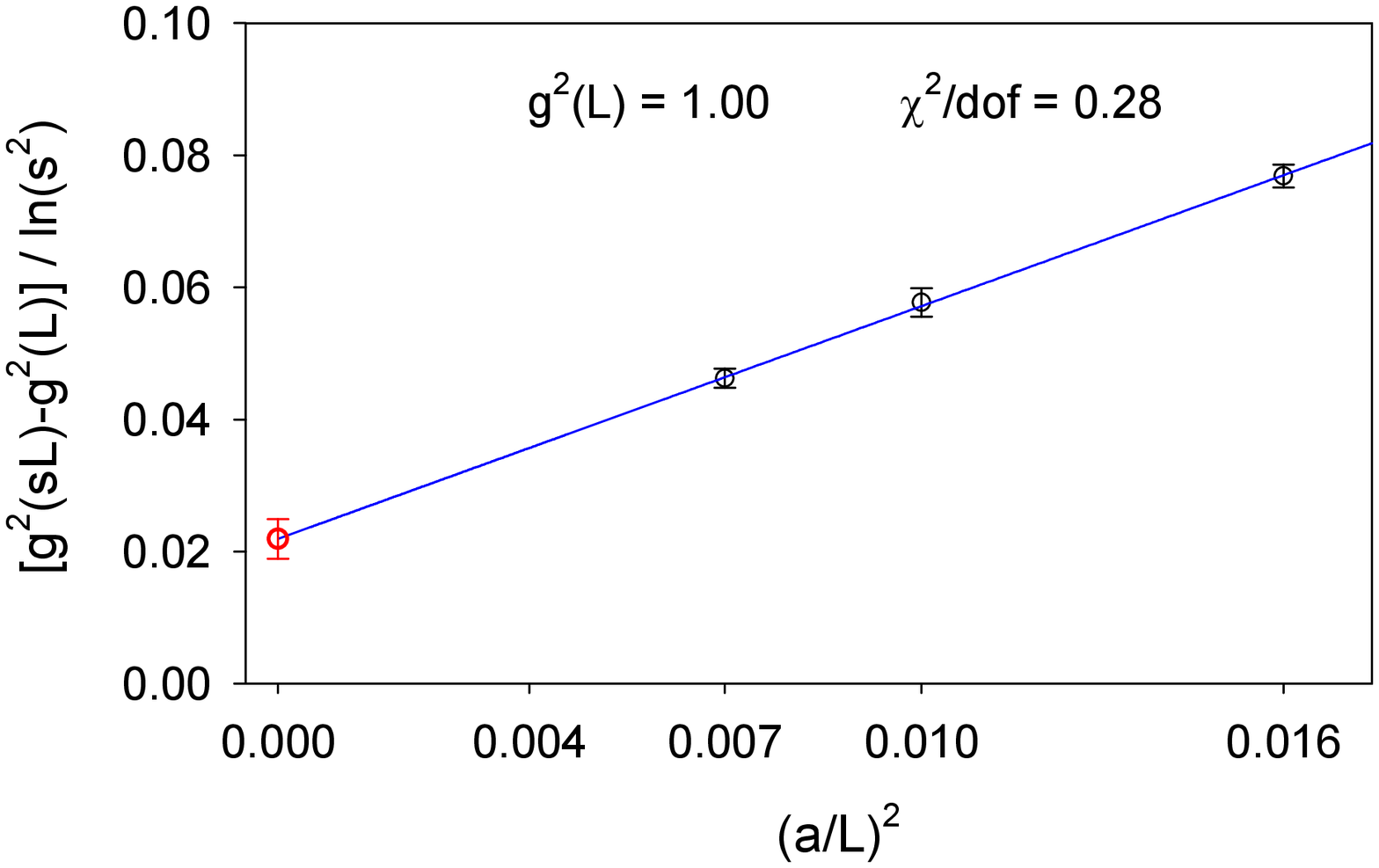} 
\\
\includegraphics*[height=5cm,width=7.5cm,clip=true]{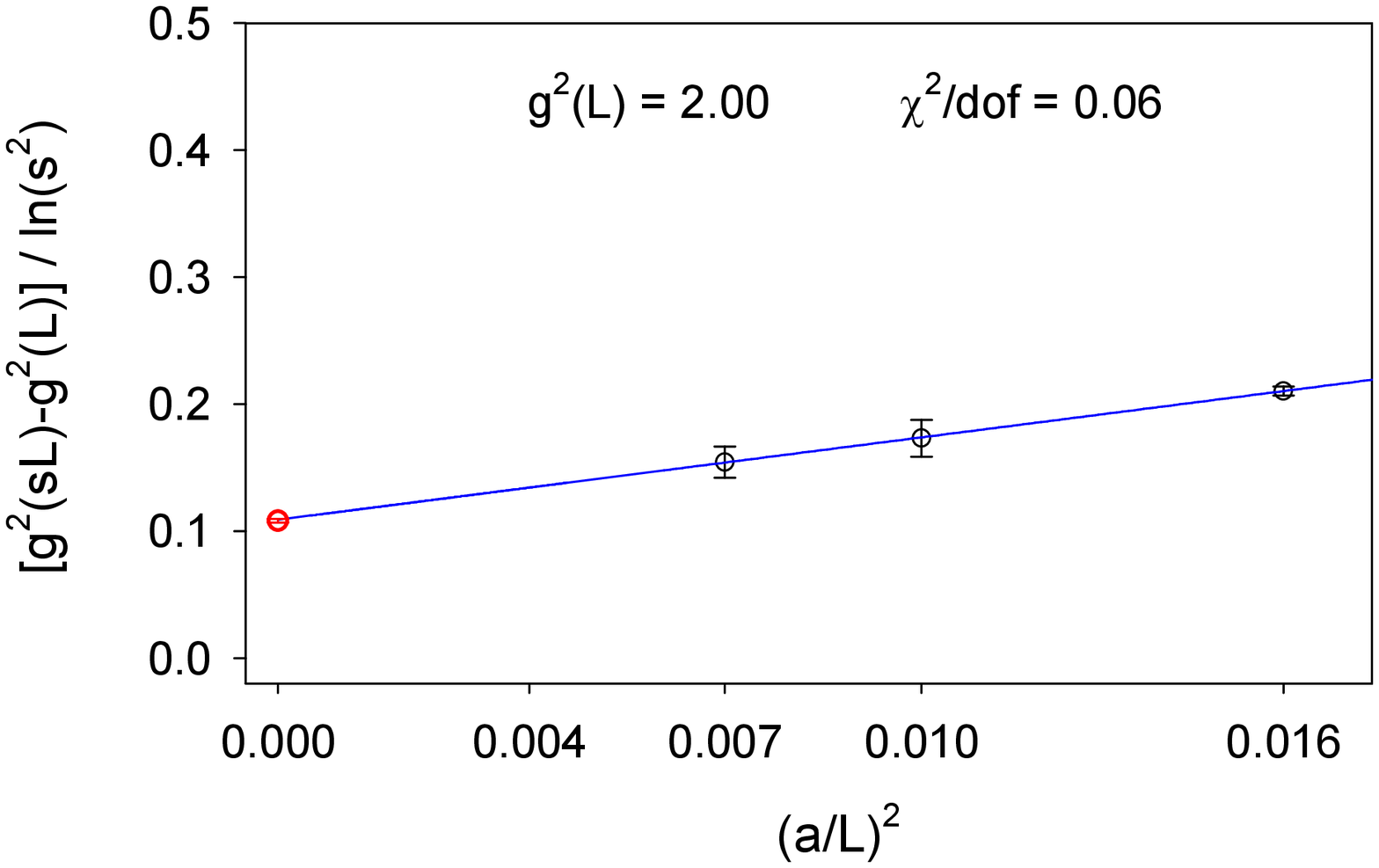}
&
\includegraphics*[height=5cm,width=7.5cm,clip=true]{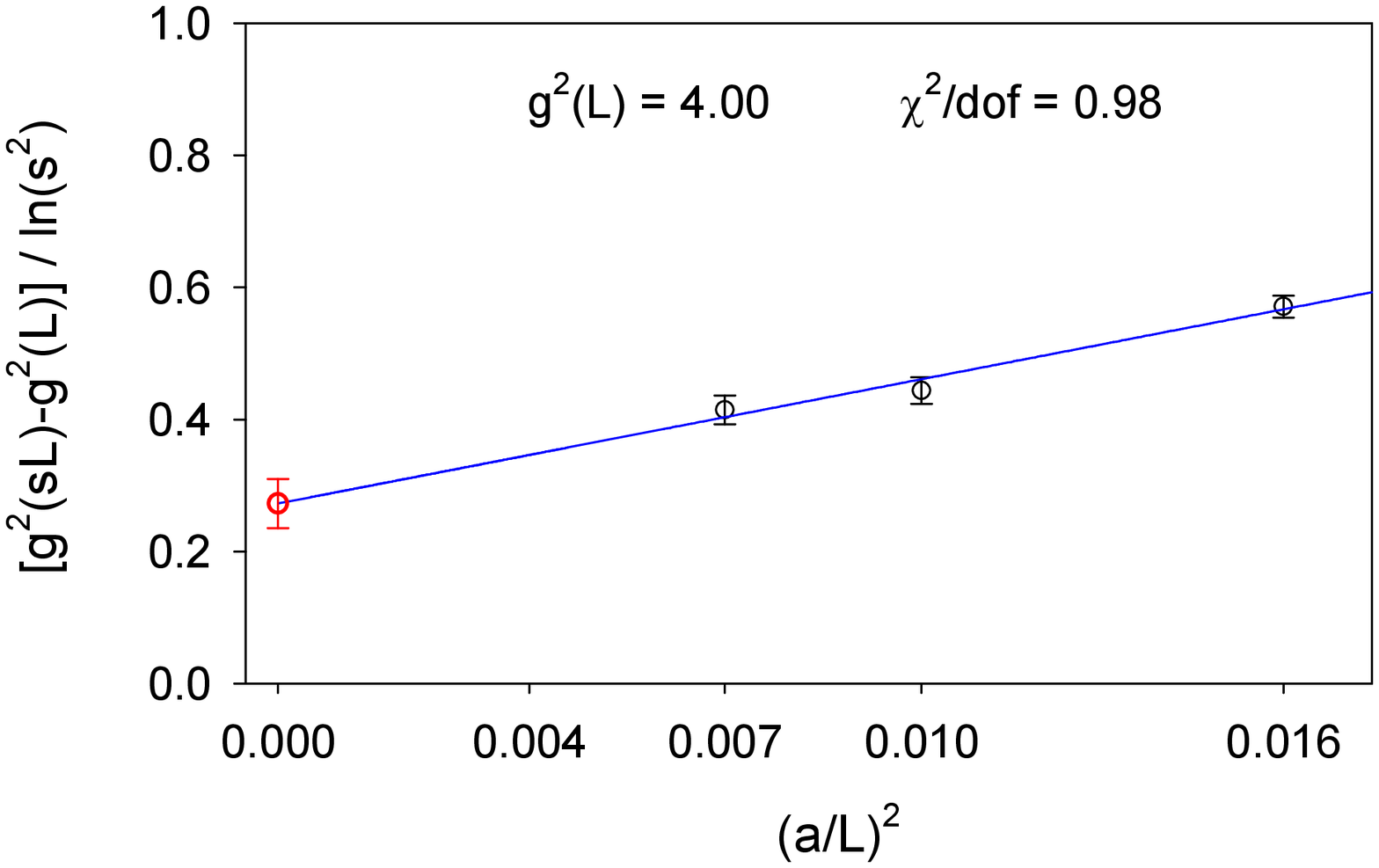} 
\\
\includegraphics*[height=5cm,width=7.5cm,clip=true]{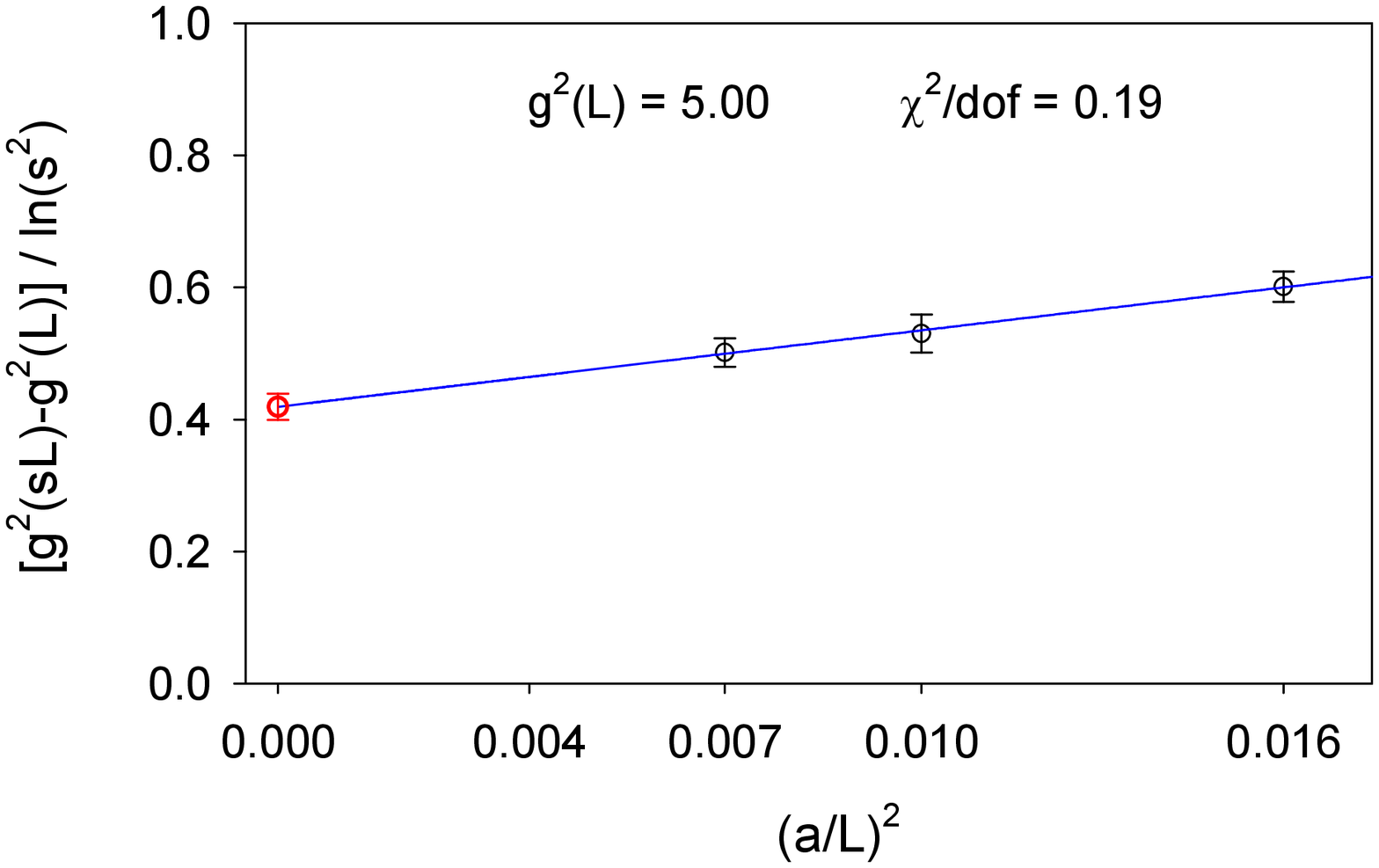}
&
\includegraphics*[height=5cm,width=7.5cm,clip=true]{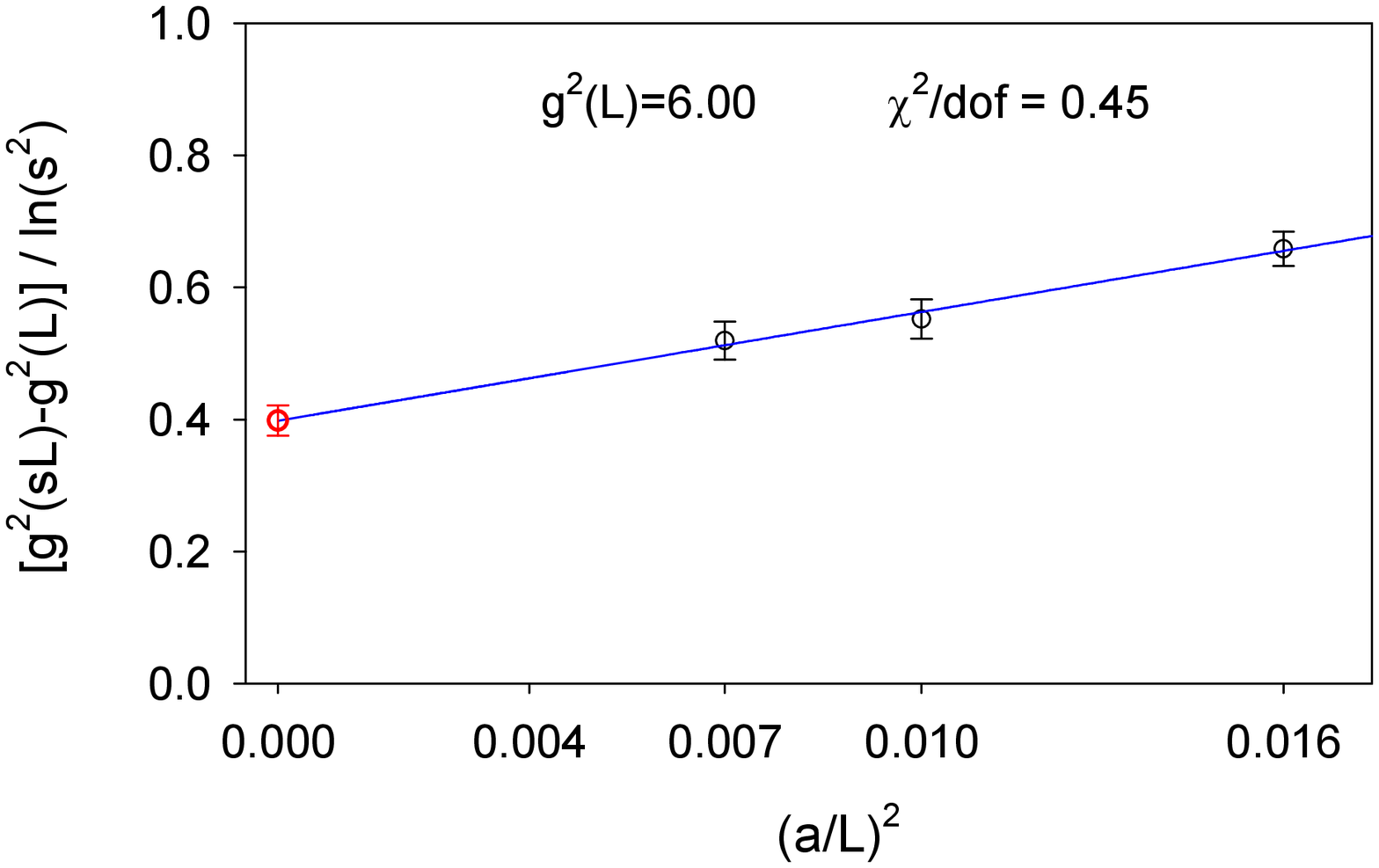}
\\
\includegraphics*[height=5cm,width=7.5cm,clip=true]{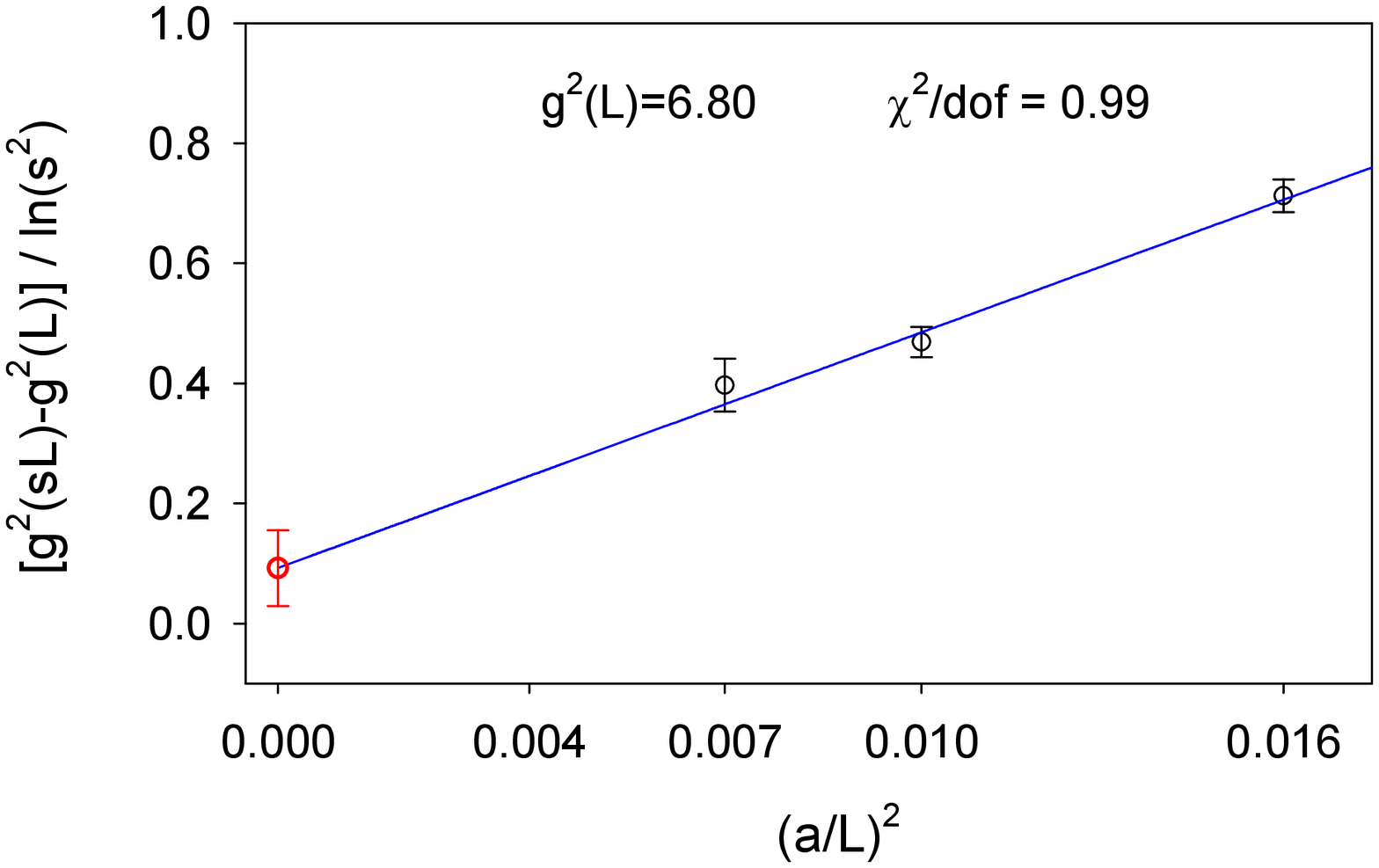}
&
\includegraphics*[height=5cm,width=7.5cm,clip=true]{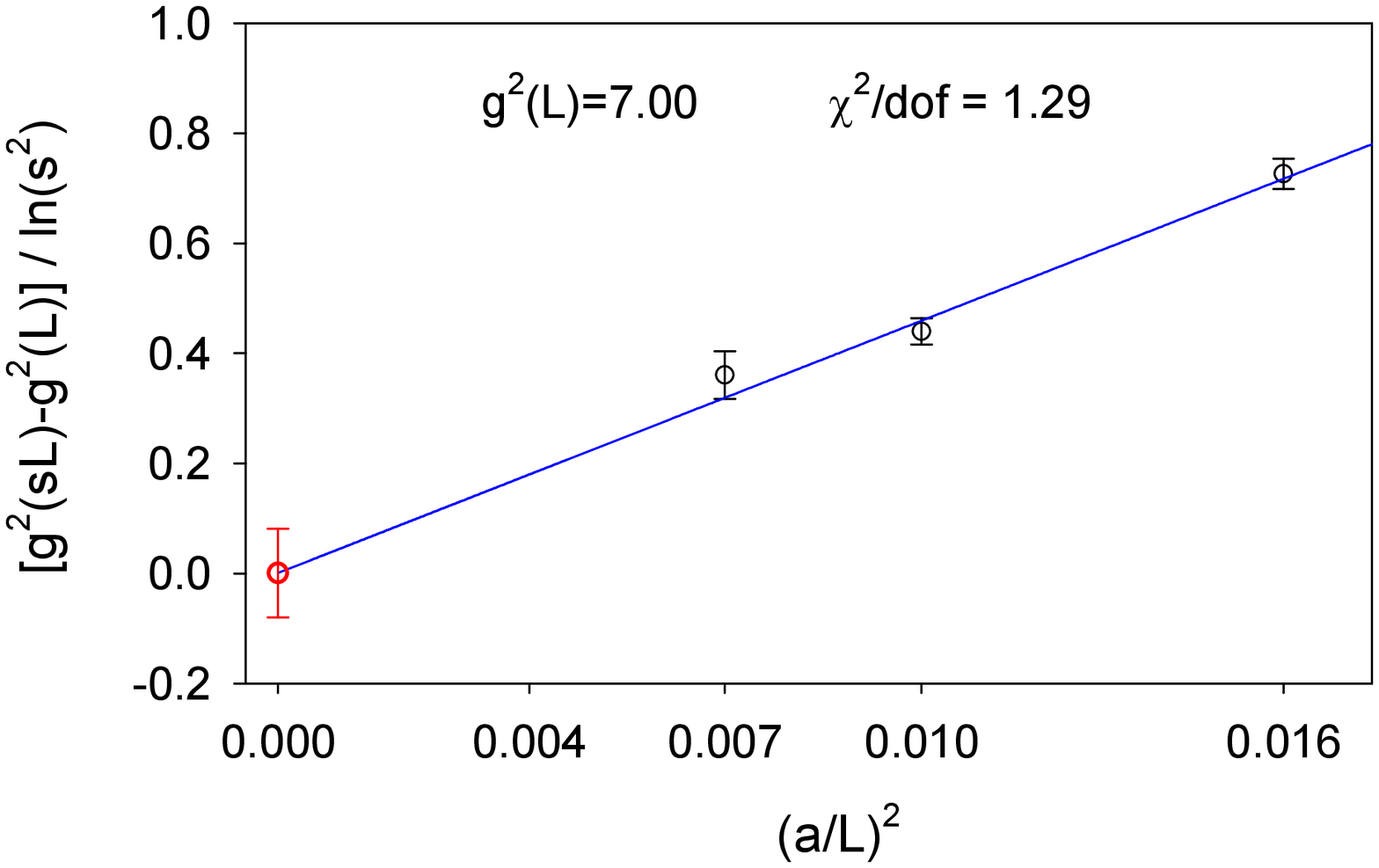}
\\
%\\ (a) & (b)
\end{tabular}
\caption{The discrete beta-functions  
of three lattice pairs $ (L, sL)/a = \{ (8, 16) $, $(10, 20)$, $(12, 24) \} $ are plotted versus $ (a/L)^2 $,   
for $ g^2(L) $ = 0.7, 1.0, 2.0, 4.0, 5.0, 6.0, 6.8, and 7.0.    
The extrapolation to the continuum limit ($ a \to 0 $) is obtained by linear fit, as shown by the straight line 
in each diagram. 
}
\label{fig:DBF_s2_g2L}
\end{center}
\end{figure}

In Fig. \ref{fig:DBF_s2_g2L}, we plot $ -\beta(2, a/L, g^2) $ 
versus $ (a/L)^2 $, for $ g^2(L) $ = 0.7, 1.0, 2.0, 4.0, 5.0, 6.0, 6.8, and 7.0, together with the extrapolation
to the continuum limit by the linear fit.
For $ g^2(L) \le 6.80 $, the data points are well fitted by a straight line with $\chi^2 /{\rm d.o.f} < 1 $.
At $ g^2(L) = 6.80 $, the linear fit gives $ -\beta(2,g^2) = 0.09(6) $ with $ \chi^2/{\rm d.o.f.} = 0.99 $,
which is quite close to the IRFP.
At $ g^2(L) = 6.90 $, the linear fit gives $ -\beta(2,g^2) = 0.03(8) $ with $ \chi^2/{\rm d.o.f.} = 1.27 $,
which is closer to the IRFP than $ g^2(L) = 6.80 $. 
At $ g^2(L) = 7.0 $, the linear fit gives $ \beta(2,g^2) = 0.00(8) $
with $ \chi^2/{\rm d.o.f.} = 1.29 $, which is consistent with zero up to the estimated error.
This suggests that $ g^2(L) \simeq 7.0 $ is an IRFP of the discrete $\beta$-function of the $SU(3)$ gauge theory
with $N_f=10 $ massless fermions in the fundamental representation,
in the finite-volume gradient flow scheme with $ c = \sqrt{8t}/L = 0.3 $.
Since the largest value of $ g^2(L,a) $ for $ L/a = 8 $ is $ 7.11(3) $,
we cannot obtain the discrete $\beta$-function for $ g^2 > 7 $.

\begin{figure}[H]
\begin{center}
\includegraphics*[width=10cm,clip=true]{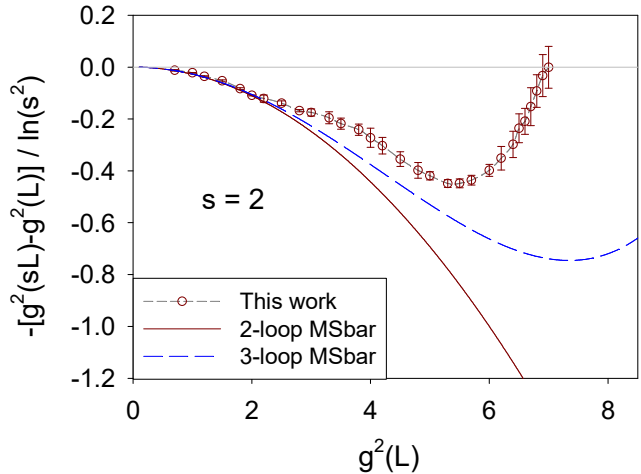} 
\caption{The discrete $\beta$-function in the continuum limit $ \beta(2, g^2) $ versus $ g^2 $,   
for the $ SU(3) $ lattice gauge theory with $ N_f = 10 $ massless optimal DWF. 
The solid and dashed lines are the 2-loop and 3-loop discrete $\beta$-functions in the $\overline{{\rm MS}} $ scheme.
}
\label{fig:DBF_s2_Nf10_ODWF}
\end{center}
\end{figure}

Our results of the discrete $\beta$-function 
in the continuum limit $ \beta(s, g^2) $ are plotted in Fig. \ref{fig:DBF_s2_Nf10_ODWF},  
together with the 2-loop and 3-loop discrete $\beta$-functions in the $\overline{{\rm MS}} $ scheme,  
\BAN
& & \beta(s,g^2(L)) = -\frac{g^2(sL) - g^2(L)}{\ln(s^2)}  \\
&=& -b_1 \frac{g^4(L)}{(4\pi)^2} - (b_1^2 \ln(s^2) + b_2) \frac{g^6(L)}{(4 \pi)^4}  
     -\left[b_1^3 (\ln(s^2))^2 + \frac{5}{2} b_1 b_2 \ln(s^2) + b_3 \right] \frac{g^8(L)}{(4 \pi)^6} + O(g^{10}),    
\EAN
where $ b_1 = 11 - 2 N_f/3 $ \cite{Gross:1973id,Politzer:1973fx}, 
$ b_2 = 102 - 38 N_f /3 $ \cite{Caswell:1974gg,Jones:1974mm}, 
and $ b_3 =  2857/2 - 5033 N_f/18 + 325 N_f^2 /54 $ \cite{Tarasov:1980au,Larin:1993tp}.

The salient features of the discrete $\beta$-functions in Fig. \ref{fig:DBF_s2_Nf10_ODWF} 
can be summarized as follows. 
In the weak-coupling regime $ g^2(L) \le 2.2 $, the lattice $\beta$-function is in good agreement 
with the 2-loop and 3-loop $\overline{\rm MS} $ results, as a decreasing function of $g^2(L) $.  
Then, in the regime $ 2.2 \le g^2(L) \le 5.3 $, the lattice, the 2-loop, and the 3-loop $\beta$-functions 
are all decreasing functions of $g^2(L)$, but with different rates. 
In the strong-coupling regime, $ 5.3 \le g^2(L) \le 7.0 $, 
the lattice $\beta$-function turns into an increasing function of $ g^2(L) $ at $ g^2(L) \simeq 5.3 $, 
and it finally reaches IRFP at $ g^2(L) \simeq 7.0 $, 
while the 2-loop and 3-loop $\beta$-functions remain as decreasing functions of $ g^2(L) $ with differnt rates.  
Note that the 3-loop $\beta$-function bends up to become an increasing function at $ g^2(L) \simeq 7.6 $,  
then attains its IRFP at $ g^2(L) \simeq 10.6 $ (out of the scale in Fig. \ref{fig:DBF_s2_Nf10_ODWF}).

\section{Discussion and Outlook}

In this paper, we perform the first study of the discrete $\beta$-function of $ SU(3) $ lattice 
gauge theory with $N_f = 10$ massless domain-wall fermions in the fundamental representation. 
Gauge ensembles are generated for 6 lattice sizes $ (L/a)^4 = 8^4, 10^4, 12^4, 16^4, 20^4, 24^4 $, 
each of 12 values of $ \beta = 6/g_0^2 \in [6.45, 15.0] $. 
The renormalized coupling is obtained by the finite-volume gradient flow scheme with $ c = \sqrt{8t}/L = 0.3 $, 
and the discrete $\beta$-function $ \beta(2, a/L, g^2) $ is extrapolated to the continuum limit 
by the step-scaling method. 
Our result of the discrete $\beta$-function $ \beta(2, g^2) $ (in Fig. \ref{fig:DBF_s2_Nf10_ODWF})
suggests that this theory possesses an infrared fixed point around $ g^2 \sim 7.0 $, 
and the theory is infrared conformal.

Our next step is to increase the statistics of the gauge ensembles with $ \beta \in [6.45, 6.60] $, 
as well as to simulate more gauge ensembles with $ \beta $ in this interval, which are now in progress. 
This will reduce the statistical and systematic errors of $ g^2(L,a) $ 
in the strong-coupling regime $ 6.0 \le g^2(L,a) \le 7.5 $, and give a more precise determination 
of the discrete $ \beta$-function and its IRFP. These results will be reported elsewhere.    

So far, we have used three lattice spacings to extrapolate $ \beta(s, a/L, g^2) $ 
to the continuum limit ($ a \to 0 $). 
It is instructive to add one more data point with finer lattice spacing 
for the extrapolation to the continuum limit. 
To this end, we are performing simulations of this theory 
%($SU(3)$ gauge theory with $N_f=10 $ massless fermions in the fundamental representation) 
on the $ 32^4 $ lattice, which will take much longer         
time than those (on smaller lattices) in this work. When the data of the $ 32^4 $ lattice 
will be ready, we can obtain the discrete $\beta$-function of 4 lattice spacings corresponding to 
4 lattice pairs $(L, 2L)/a$ = (8, 16), (10, 20), (12, 24), and (16, 32).
These results will be reported in a forthcoming publication.

In this paper, we have used the optimal DWF with $R_5$ symmetry \cite{Chiu:2015sea} for the massless fermions. 
Nevertheless, we are planning to repeat the same study with the original  
optimal DWF without $ R_5 $ symmetry \cite{Chiu:2002ir}, which has better chiral symmetry. 
It is interesting to compare results obtained with these two slightly different formulations 
of lattice fermion, especially for the strong-coupling regime, $ 6.42 \le \beta  \le 6.60 $.  

Moreover, we are extending our present study to the cases of $ N_f = 8 $ and $ N_f = 12 $ 
massless domain-wall fermions. These studies are required for the determination of the conformal window 
of $ SU(3) $ gauge theory with $ N_f $ massless fermions in the fundamental representation.

\begin{acknowledgments}

  Part of this work was performed during a visit at the University of Wuppertal.
  The author is grateful to Zoltan Fodor and Wuppertal lattice group for kind hospitality and support, 
  as well as useful discussions. 
  The author thanks Chik-Him Wong and the authors of Ref. \cite{Fodor:2014cpa} for providing the numerical data 
  of $ \delta(c,a/L) $. 
  The author would like to thank the organizers of the workshop  
  ``Lattice Gauge Theory Simulations Beyond the Standard Model of Particle Physics"  
  held at Tel Aviv University in June 2015. 
  The author is indebted to Yu-Chih Chen, Han-Yi Chou, and Tung-Han Hsieh for their help in the code development. 
  This work is supported by the Ministry of Science and Technology  
  (No.~NSC102-2112-M-002-019-MY3), Center for Quantum Science and Engineering 
  (Nos.~NTU-ERP-103R891404, NTU-ERP-104R891404, NTU-ERP-105R891404), 
  and National Center for High-Performance Computing.  

\end{acknowledgments}

%\eject

\end{document}